# The relationship between the interdisciplinary activation of children's scientific concepts and their mastery of basic knowledges：

# A pre-study based on reaction times

WANG Zhong; CUI Zhen; ZHANG Yi

*(Beijing Doers Education Consulting Co.,Ltd; Beijing Dongcheng District Qianmen Primary School ;Beijing Fengtai District Fengtai First Primary School)*

**Abstract:** the activation of scientific concepts (such as association) is not only an important way for children to organize scientific knowledges, but also an important way for them to learn complex concepts (such as compound concepts composed of multiple knowledge points). Inspired by the details of a primary electrical lesson, we used the E-Prime software to study the activation of concepts inside and outside the electrical unit by students with different electrical knowledge levels (taking reaction time as the main index). The results showed that: ① the levels of basic knowledge was negatively correlated with the cross domain activation ability of concepts, that is, the worse the basic knowledge, the faster the cross domain activation speed (P < 0.05); ② The better the basic knowledges, the closer the activation behavior is to science teachers. Conclusion: ① the poorer the basic knowledges, the stronger the cross domain association ability and the more active the thinking; ② The reason for this phenomenon may be human's thinking is patterned; ③ Therefore, there is a potential logical contradiction in the view that divergent thinking is the psychological basis of creativity.

One concept activates another (such as associating from one concept to another), which belongs to the organizational mode of knowledge representation, and it reflects the way of human brain to deal with a large amount of knowledge. And the activation of concepts such as association will produce new learning results: "association learning is an effect of individual behavior changes caused by the connection between different stimuli or events in the environment"(LV Xiaojing, Ren Xuezhu, 2018). Therefore, the activation of concepts plays an important role in children's learning. However, what factors affect the activation and invocation of such concepts? What is the main resistance? What is the "farthest transmission power" and "coverage" of this activation capability?

## 1. Origin of this study

The research of this paper originates from a scientific lesson about electrical. In the first half of 2022, one of the authors of this paper reported the details of a lesson: in a class of "conductors and insulators", the teacher guided the students to explore the conductivity of the objects at hand through experiments, and asked them to divide these objects into "conductors" or "insulators". In the middle of the course, the teacher suddenly asked the students: "is the magnet a conductor?", and guide students to experiment and classify the magnet.

The uniqueness of this lesson detail is not whether the magnet is a conductor (at present, most of the magnet used in primary schools are made of ferromagnetic materials, so the magnet in this lesson is naturally a conductor), but "magnet" is not a concept of electrical unit. Take the textbook of Educational Science Publishing House,China (which was used in this lesson) as an example, the "magnet" belongs to the "MAGNET" unit of Grade 2, while the "conductor and insulator" belongs to the "CIRCUIT" unit of Grade 4 (Yu bo, 2020). Generally speaking, even if teachers see the magnet, it is difficult for them to associate with it in the direction of "conducting electricity", and they will not guide students to try so. (we conducted a survey on this issue among science teachers in Tongzhou District of Beijing by using the APP of "questionnaire star". Among the 57 valid questionnaires, only 16 mentioned that the magnet was involved in the last teach about "conductors and insulators", accounting for 28.1% of the valid questionnaires).

We investigated the background of the teacher, and learned that he was a post transfer teacher (i.e. the major he studied and the discipline he initially taught was not the subject of science), and he's basic knowledges of this discipline was relatively weak. From this, we wonder whether the teacher's active imagination is related to his weak basic knowledges? In other words, it is the weakness of basic knowledges that make it easier for him to activate cross domain (different units) concepts without so many "rules and regulations" in his mind. Is that led him linked "conductivity" with "magnet"?

## 2. Reviews

The activation of concepts belongs to the representation and organization of knowledges. Previous studies have shown that the activation of concepts is not random, but has its internal laws.

For example, Collins and Quillian found in 1969 that people's reaction time to infer a sentence that spans two semantic levels (such as "canary is a bird") is shorter than three semantic levels (such as "canary is an animal"). For another example, rips et al found in 1973 that people always respond more quickly to the sentence "robin is a bird" than "turkey is a bird". Rips et al believe that this is because "for most people, robin is a typical bird, but turkey is not" (Kathleen M. Galotti, 2017, p113).

Further studies found that the difference in activation was related to people's organizational law of concepts. For example, Wu Yanan and others proposed that children have two basic preferences for the organization of concepts: taxonomic relationship and thematic relationship. Taxonomic relationship is a taxonomic division of things / objects based on their inherent shared characteristics. For example, "pig" and "sheep" belong to the category of "mammal". And thematic relationship refers to the external relationship between things / objects in the same scene or event. For example, "Jungle" and "bird", "blackboard" and "chalk" all belong to the same scene (Wu Yanan et al., 2019). Wu Yanan and others believe that these two relationships are the two modes that children mainly follow in conceptual organization and activation. However, is there a master-slave relationship or competition between these two relationships? In response, Bi Yanchao and others used the semantic concepts of "doctor" and "teacher" (taxonomic relationship) and the semantic concepts of "doctor" and "stethoscope" (thematic relationship) as stimulation conditions, and used functional nuclear magnetic resonance (fMRI) to study the brain oxygen dependent signals of the participants. It was found that taxonomic representation was the main organizational dimension of neural representation, while thematic representation was embedded in but independent of taxonomic representation (Yangwen Xu, Xiaosha Wang, Xiaoying Wang et al, 2018).

All these studies show that people's organization and activation of concepts not only have complex structures and laws, but also the structures have strict hierarchy. Then, if we regard the two relationships as two domains, it is obvious that cross domain activation of concepts is not easy. However, this is only our inference from the literature reviews. The specific entry point of this paper is not a hot point in previous studies.

In terms of research methods, it has been proved to be an effective method to use reaction time to measure the level of participants' conceptual activation. For example, Meyer and schvaneveldt used the spreading activation model in 1971 to study the conceptual activation of the participants. They found that if a letter string is a real word (such as "bread"), participants will respond more quickly to words that are semantically related to the word (such as "butter") than an unrelated word (such as "chair") or a nonexistent word (such as "rencle") (Kathleen M. Galotti, 2017, p112). In addition, Zhang Jijia and others have studied the relationship between the cultural differences of different ethnic minorities in China (Yi, Bai and Mosuo) and the activation of kinship vocabulary. That research adopted the reaction time method, "the reaction time of the elder kinship word pair, 'upper male and lower female' word pair is significantly shorter than that of 'upper female and lower male'" (Zhang Jijia, 2020).

To sum up, the cross domain activation of children's scientific concepts should not be arbitrarily, and their basic knowledges may be an important influence factor (may be a stumbling block). On the other hand, using reaction time to test the concept activation has been proved to be a feasible method.

## 3. The basic conjecture of this study

Children's cross domain activation of scientific concepts may be affected by their basic knowledge levels and may be negatively correlated, that is, the worse the level of basic knowledge, the stronger their cross domain association ability and the faster their reaction speed.

## 4. Experimental designed

### 4.1 Basic idea

We use the pupils in one class with medium learning achievements and good intelligence to conduct basic knowledge test first, with the purpose of selecting the students with the highest and lowest scores and listing them in the "high score group" and "low score group" respectively. The selection criteria was the scores of the two groups should be significantly different in statistic (using independent sample t-test, the standard was $p < 0.05$). Then, as an experimental stimulus, the two groups were asked to complete the same set of questions containing both internal and external concepts of the electrical unit (represents in and across the domain) to obtain the reaction time data of the two groups of participants. Finally, the reaction time data of the two groups were processed to test whether there were significant differences between them (using independent sample t-test, the standard was $p < 0.05$), and the conclusion of this paper was reached finally.

4.2 Experimental process:

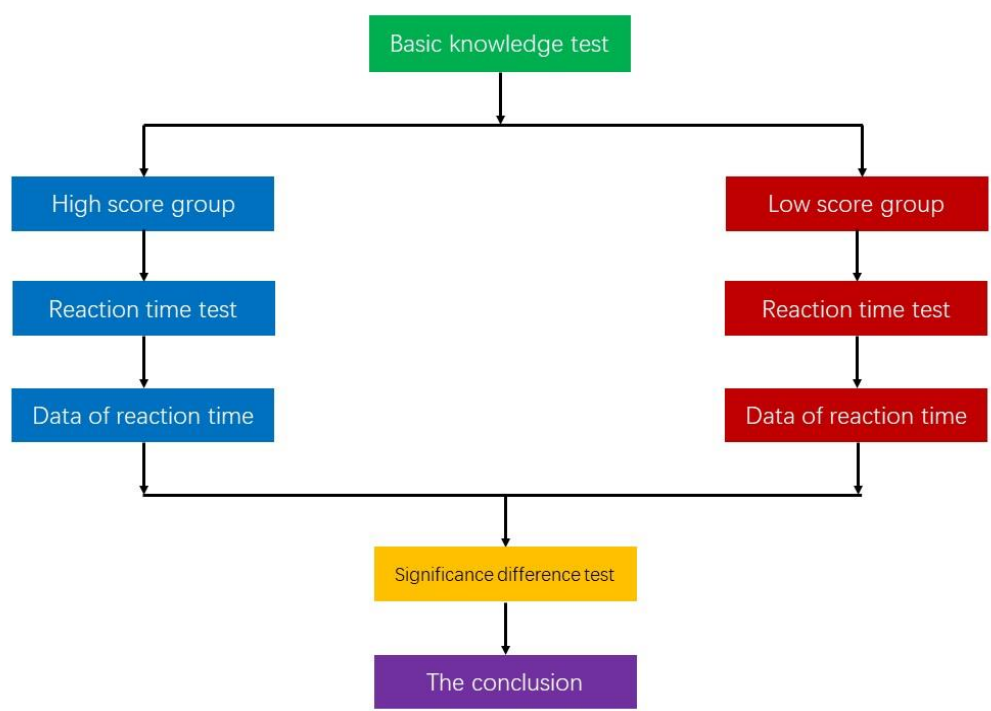

Figure 1

4.3 Concept points and interrelations contained in electrical units:

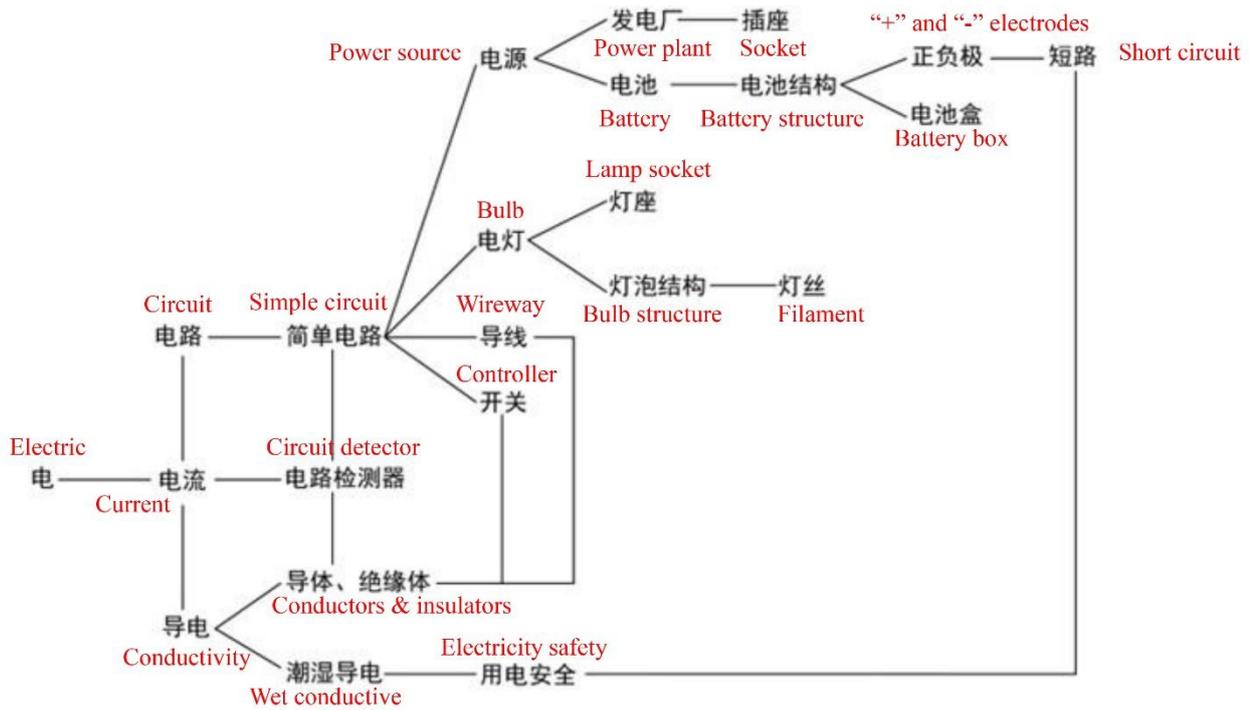

Figure 2

Textbook version: Educational Science Publishing House,China (Yu Bo,2020).

4.4 Basic knowledge test questions: see "10.1 the basic knowledge test" for details.

Note: in order to ensure a certain degree of discrimination, the basic knowledge test questions are slightly more difficult.

4.5 Reaction time questions

| Serial number | Questions of the reaction time task | Whether out unit? | The concepts out of the unit |
|---|---|---|---|
| Q1 | The battery has north and south poles and must not be reversed | N | —— |
| Q2 | Simple circuits generally do not contain electromagnets | Y | electromagnet |
| Q3 | A magnet is a conductor | Y | magnet |
| Q4 | There are both conductors and insulators in the bulb | N | —— |
| Q5 | A circuit is a closed loop | N | —— |
| Q6 | The short circuit is caused by water | N | —— |
| Q7 | A battery is the power source | N | —— |
| Q8 | If a computer is broken, probably because the circuit | Y | computer |
| | inside is not closed | | |
| Q9 | Purified water can conducts electricity | N | —— |
| Q10 | A simple circuit is like a "circle" | N | —— |
| Q11 | The filament can conduct electricity | N | —— |
| Q12 | There are closed circuits in an electronic watch | Y | electronic watch |
| Q13 | There is no insulator in the circuit because it can conductive | N | —— |
| Q14 | When you send a message, an electric current flows through your cell phone | Y | cell phone |
| Q15 | The circuit generally contains batteries | N | —— |

Table 1

See "4.3" above for whether the concept of the test questions is cross unit.

## 5. Experiment preparation and Implementation

5.1 Participants preparation

We selected 22 students from a fifth-grade class of a primary school in Dongcheng District of Beijing as the participants. At the same time, one of the authors, as an adult participant, was also collected reaction time data for comparison. The reason why fifth-grade is chosen instead of fourth-grade is that the fourth-grade learned the electrical knowledges online, not at school (because of the epidemic of Beijing), so the learning effect is difficult to guarantee.

5.2 Instrument preparation:

E-Prime 2.0 was selected as the reaction time data acquisition software and SPSS 24.0 was selected as the data processing software. The stimulation presentation instrument was a REDMI pro14 Ryzen laptop (14 inches, 60.008 Hz).

5.3 The procedure of reaction time task

The stimulation of E-Prime software is divided into two procedures: exercises procedure and formal experiment procedure, and these two procedures need to be started respectively during the experiment. The former is to familiarize the participants with the use of the software. The programming ideas of the two are completely the same, and the difference in stimulus materials is the only difference between the two.

First of all, the software will present the experimental instruction as a picture. After being understood by the participants, press the space bar to enter the stimulation task. At the first of the stimulation task is, the red fixation point "+" will appear in the middle of the screen for 250ms. Then the question number "question X" will be presented for 250ms, and then the stimulation materials will be presented. The stimulation materials for the exercise procedure are common sense questions like "you are a boy" or "you are a fourth-grade student". See section "4.5" for the stimulation materials for the formal experiment. There are 20 exercises and 15 formal experiment questions. All questions are presented in black letters on a white background, Chinese in bold letters (No. 36), all English letters and figures in Times New Roman letters (No. 36). After judging whether it is right or not, the participants need press J or F to answer and enter the next question (correct press J, wrong press F). The software will automatically jump to the next question after waiting for 5000ms for no answer. After the test, the screen displays the closing words: "the experiment is over, thank you for your participation!". After 3000ms, the program pops out and returns to the E-Prime 2.0 editing interface.

The procedure flow is such as the following figure.

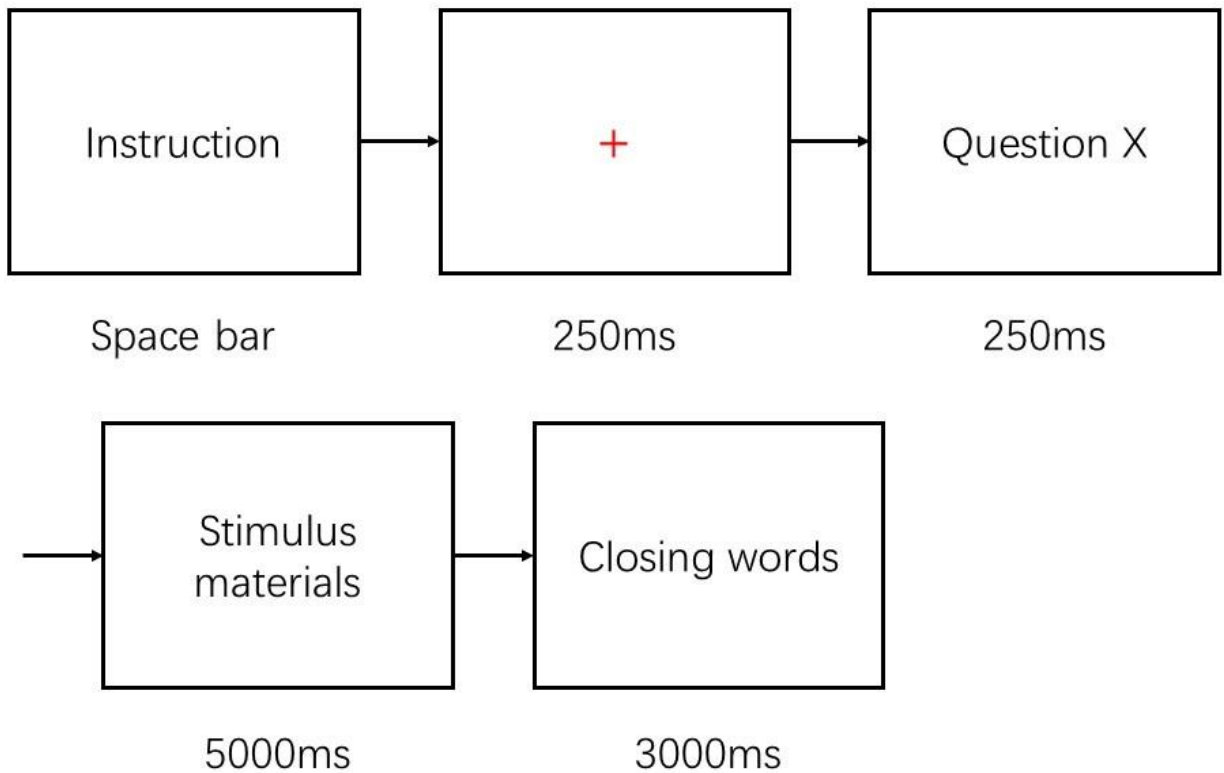

Figure 3

5.4 The work to prevent guess

In order to prevent the participants from guess the purpose of the experiment, we have done the following works:

① The stimulus materials of the reaction time test included not only the concepts within the unit, but also the concepts out of the unit (even the concepts out of the science textbook of primary schools, such as cell phone).

② Only told the participants about the test may involve electrical knowledges, but not the specific details and the purpose of the experiment;

③ All the 22 students were not informed of the existence of another test (reaction time task) when they participated in the basic knowledge test;

④ Since all the participants in the reaction time task came from the same class and know each other, they was likely to guess the experimental purpose according to the previous learning achievements of each other(They may guess that there are 2 groups of "high score" and "low score" in the experiment). Therefore, during the reaction time task, we disordered the order of high score and low score groups and randomly select them to complete the task, making it difficult for them to guess.

⑤ Although the adult participant knew the general purpose of the experiment, he had not seen the questions of the reaction time task.

The interview after the reaction time test showed that all participants (except adults) could not accurately state the purpose of the experiment.

## 6. Results and analysis

6.1 About the basic knowledge test

Among all 22 students, we selected 3 students with the highest scores and 3 students with the lowest scores as the reaction time participants, accounting for 27.27% of all students.

The following tables shows the group statistics and independent sample test of the basic knowledge test (Tables 2 and 3). The effect size measured by Hedge's g value, is 8.85, which is extremely significant (Cohen's method is used to calculate the sample effect size in this paper (c). If the sample number is less than 10, Hedges' g method is used (d), the same below). According to table 3, the comparison between the high score group and the low score group conforms to the homogeneity test, and the statistical difference is extremely significant (significance (two tailed) is 0.000, less than 0.01), meeting the selection criteria.

| | Number of samples | Average | SD | SEM |
|---|---|---|---|---|
| High score group | 3 | 82.6667 | 10.06645 | 5.81187 |
| Low score group | 3 | 13.3333 | 4.61880 | 2.66667 |

Table 2

| | Levene test | | | | Independent t-test | | | | |
|---|---|---|---|---|---|---|---|---|---|
| | F | Significance | t | df | Significance (two tailed) | Average interpolation | SE interplation | 95% CI of interpolated | |
| | | | | | | | | Lower limit | Upper limit |
| EV* | 1.362 | 0.308 | 10.843 | 4 | 0.000 | 69.33333 | 6.39444 | 51.57952 | 87.08715 |
| Un-EV | | | 10.843 | 2.806 | 0.002 | 69.33333 | 6.39444 | 48.16550 | 90.50116 |

Table 3

*"EV" means the equal variance. Abbreviations are used because of insufficient table space, and the sames below*

## 6.2 About total reaction times

"10.2 the data of total reaction times" shows the total reaction times and average correct rates of the 7 participants.

Table 4 and table 5 show the group statistics and independent sample test results of the total reaction time of the 6 participants. The effect size of the data is extremely significant (c = 0.91). It easy to see that the reaction times of the high score group is significantly greater than that of the low, and the statistical difference of the data is extremely significant (P = 0.000). It reflects that the activation of concepts of the low score group is easier and the thinking is more flexible.

| | Number of samples | Average | SD | SEM |
|---|---|---|---|---|
| High score group | 45 | 2829.2667 | 846.61297 | 126.20561 |
| Low score group | 45 | 2124.3111 | 719.63870 | 107.27740 |

Table 4

| | Levene test | | Independent t-test | | | | | | |
| | F | Significance | t | df | Significance (two tailed) | Average interpolation | SE interplation | 95% CI of interpolated | |
| | | | | | | | | Lower limit | Upper limit |
| EV | 1.327 | 0.252 | 4.256 | 88 | 0.000 | 704.95556 | 165.63906 | 375.78274 | 1034.12837 |
| Un-EV | | | 4.256 | 85.774 | 0.000 | 704.95556 | 165.63906 | 375.66366 | 1034.24746 |

Table 5

## 6.3 Reaction times about cross domain concepts

But is the conclusion above an illusion? We special selected and analyzed the reaction times of all five cross domain activation questions separately (see "10.3 the data of reaction times about cross domain concepts" for details)

Table 6 and table 7 show the group statistics and independent sample test results under this dimension. The effect size of the data is extremely significant (c = 1.07). When completing the crossing domain activation task, the reaction times of the high score group is still significantly slower than that of the low, and the statistical difference of the data is extremely significant too (P = 0.008). This not only confirms the results of the above total reaction time data, but also is the most direct key evidence of this study.

| | Number of samples | Average | SD | SEM |
| --- | --- | --- | --- | --- |
| High score group | 15 | 2932.7333 | 893.53379 | 230.70943 |
| Low score group | 15 | 2153.2000 | 581.38777 | 150.11368 |

Table 6

| | Levene test | | Independent t-test | | | | | | |
| | F | Significance | t | df | Significance (two tailed) | Average interpolation | SE interplation | 95% CI of interpolated | |
| | | | | | | | | Lower limit | Upper limit |
| EV | 2.102 | 0.158 | 2.832 | 28 | 0.008 | 779.53333 | 275.24709 | 215.71524 | 1343.35143 |
| Un-EV | | | 2.832 | 24.052 | 0.009 | 779.53333 | 275.24709 | 211.51670 | 1347.54997 |

6.4 Eliminate the possibility interference that the habit of reading questions

However, there is still a problem left here: generally speaking, students with good achievements have a good habit of reading the questions carefully. Then, will the slower reaction time of the participants in the high group be related to their habit of reading the questions (that is, they spend more time on observe and thinking the questions)? In another word, is the good habit a possibility interference of this study?

We believe that the possibility of this factor is very small for the following reasons:

6.4.1 Average accuracy

It can be seen from the data of total reaction times that the average correct rate of the high score group is lower than that of the low score group (0.49 for the high and 0.67 for the low). That is to say, suppose the extra time is spent on reading and understanding the questions, but it still does not more helpful of accuracy. Considering that all the participants in this group are students with good basic knowledge level, this is even more contradictory.

6.4.2 Variances comparison of reaction times

Suppose more time is spent on reading the questions, then the reaction behaviors of high score group participants should be different from that of the low participants because they obviously need more inspect time. In that case, their reaction time variance should be significantly different from that of the low score group, so as to show the existence of more observation (reading) behaviours.

Previous studies have confirmed the value of this method. For example, Li Qing has studied the difference in the reaction times of mathematical excellent students and students with learning difficulties to problem representation when completing application problems. The following figure is one of the data tables in that paper, and SD is the variance ("single poor students" (单困) are students with learning difficulties only in mathematics, and "double poor students"(双困) are students with learning difficulties judged by both mathematics and Chinese). We used SPSS to calculate the statistical difference of reaction time(反应时) variance between the excellent students and the "double poor students" in this table (because the double poor students are involved in learning difficulties in reading comprehension naturally). As shown in table 8, the reaction time variances of the two showed great difference (P = 0.000) (Li Qing, 2009). The reading behaviors of people with different reading abilities is definitely different (Gao Xiaomei, 2010). Therefore, we can see the influence of reading behavior on the variance of reaction times.

表3-1-3 三类小学生各项反应正确频次及反应时

|  | 学业类型 | 数字正确频次 | 数字反应时 | 关系词正确频次 | 关系词反应时 | 变量词正确频次 | 变量词反应时 | 其它正确频次 | 其它反应时 |
|---|---|---|---|---|---|---|---|---|---|
| **N** | 学优 | 31 | 31 | 31 | 31 | 31 | 31 | 31 | 31 |
|  | 单困 | 30 | 30 | 30 | 30 | 30 | 30 | 30 | 30 |
|  | 双困 | 31 | 31 | 31 | 31 | 31 | 31 | 31 | 31 |
| **M** | 学优 | 7.48 | 257.94 | 6.94 | 271.70 | 3.13 | 306.65 | .52 | 358.77 |
|  | 单困 | 7.00 | 278.17 | 6.33 | 302.07 | 1.13 | 356.10 | .83 | 387.77 |
|  | 双困 | 6.84 | 288.29 | 5.10 | 279.58 | .48 | 324.26 | .29 | 432.54 |
| **SD** | 学优 | .724 | 50.25 | .964 | 52.63 | 1.45 | 55.30 | .81 | 51.99 |
|  | 单困 | 1.02 | 48.47 | 1.60 | 63.52 | 1.36 | 62.47 | 1.46 | 58.68 |
|  | 双困 | 1.44 | 44.87 | 2.27 | 50.85 | .68 | 64.27 | .59 | 37.50 |
| **Min** | 学优 | 6.00 | 204.00 | 5.00 | 200.00 | 1.00 | 230.00 | .00 | .63 |
|  | 单困 | 5.00 | 204.00 | .00 | 217.00 | .00 | 247.00 | .00 | 295.00 |
|  | 双困 | 3.00 | 222.00 | .00 | 210.00 | .00 | 30.00 | .00 | 301.00 |
| **Max** | 学优 | 8.00 | 417.00 | 8.00 | 435.00 | 7.00 | 408.00 | 2.00 | 472.00 |
|  | 单困 | 8.00 | 358.00 | 8.00 | 436.00 | 4.00 | 475.00 | 5.00 | 497.00 |
|  | 双困 | 8.00 | 432.00 | 8.00 | 419.00 | 2.00 | 426.00 | 2.00 | 436.00 |

Figure 4

| Levene test | | | | | Independent t-test | | | |
|---|---|---|---|---|---|---|---|---|
| F | Significance | t | df | Significance (two tailed) | Average interpolation | SE interplation | 95% CI of interpolated | |
|  |  |  |  |  |  |  | Lower limit | Upper limit |

| | | | | | | | | | |
|---|---|---|---|---|---|---|---|---|---|
| EV | 1.532 | 0.283 | 14.343 | 4 | 0.000 | 158.0033 | 11.0157 | 127.4189 | 188.5878 |
| Un-EV | | | 14.343 | 3.307 | 0.000 | 158.0033 | 11.0157 | 124.7181 | 191.2886 |

Table 8

Table 9 shows the independent test results of the reaction time variances of the two groups of participants. From the results, we can not only see that there is no significant difference between the two groups of data, but also that they are almost identical (P = 0.797). This reflects that there is not much difference in the behaviors of the two groups of participants in answering, and there is no evidence that they spend more time on reading the questions.

| | Levene test | | Independent t-test | | | | | | |
|---|---|---|---|---|---|---|---|---|---|
| | F | Significance | t | df | Significance (two tailed) | Average interpolation | SE interpolation | 95% CI of interpolated | |
| | | | | | | | | Lower limit | Upper limit |
| EV | 5.461 | 0.080 | -0.275 | 4 | 0.797 | -57.60112 | 209.15795 | -638.31668 | 523.11444 |
| Un-EV | | | -0.275 | 2.459 | 0.804 | -57.60112 | 209.15795 | -814.22216 | 699.01992 |

Table 9

## 6.5 Comparison with adult participant

As previously mentioned, we also collected data from one adult (one author of this paper) to investigate who had the activation pattern closer to that of adults in the two groups. Since the adult participant is a science teacher, and the knowledges of electricity and magnetism is the content that he needs to teach at ordinary times, we default that the basic knowledge score of the participant is full.

### 6.5.1 Total reaction times

Table 10-13 shows the group statistics and independent sample statistics of the total reaction times of high score group-adult and low score group-adult. It can be seen from the tables that, in contrast, the reaction times of the participants in the high score group is significantly closer to that of the adult (there is no significant difference between the high score group and the adult, and the difference between the low is extremely significant). It should be noted that the effect size of this part deviates: $c_{h-a}$ (c of high score group-adult) = 0.33, $c_{l-a}$ (c of low score group-adult) = 1.49. The reason for this effect size may be the amount of adult samples was too little (we only collected the data of one adult

person). Therefore, the result effect of high score group-adult should be applied with caution.

| | Number of samples | **Average** | SD | SEM |
|---|---|---|---|---|
| High score group | 45 | 2718.1556 | 882.58744 | 131.56837 |
| Adult | 15 | 2997.4000 | 749.14254 | 193.42777 |

Table 10

| | Levene test | | | | Independent t-test | | | | |
|---|---|---|---|---|---|---|---|---|---|
| | F | **Significance** | t | df | **Significance (two tailed)** | Average interpolation | SE interplation | 95% CI of interpolated | |
| | | | | | | | | Lower limit | Upper limit |
| EV | 0.035 | 0.852 | -1.099 | 58 | 0.276 | -279.24444 | 254.10430 | -787.88947 | 229.40058 |
| Un-EV | | | -1.194 | 28.042 | 0.243 | -279.24444 | 233.93276 | -758.40204 | 199.91315 |

Table 11

| | Number of samples | **Average** | SD | SEM |
|---|---|---|---|---|
| Low score group | 45 | 2013.2000 | 647.97358 | 96.59420 |
| Adult | 15 | 2997.4000 | 749.14254 | 193.42777 |

Table 12

| | Levene test | | | | Independent t-test | | | | |
|---|---|---|---|---|---|---|---|---|---|
| | F | **Significance** | t | df | **Significance (two tailed)** | Average interpolation | SE interplation | 95% CI of interpolated | |
| | | | | | | | | Lower limit | Upper limit |
| EV | 1.326 | 0.254 | -4.899 | 58 | 0.000 | -984.20000 | 200.88415 | -1386.3133 | -582.08669 |
| Un-EV | | | -4.552 | 21.429 | 0.000 | -984.20000 | 216.20532 | -1433.2757 | -535.12427 |

Table 13

### 6.5.2 Reaction times about cross domain concepts

Table 14-17 shows the statistical results of high score group-adult and low score group-adult when completing cross domain activation tasks. Consistent with 6.5.1, the high score group participants are still closer to the adult. At the same time, the effects of the two groups were extremely significant: $c_{h-a} = 1.10$, $c_{l-a} = 3.11$

In short, for the cross domains concept activation pattern, the better the basic knowledges is mastered, the closer the pattern is to the science teacher.

|  | Number of samples | Average | SD | SEM |
|---|---|---|---|---|
| High score group | 15 | 2932.7333 | 893.53379 | 230.70943 |
| Adult | 5 | 3780.0000 | 426.77453 | 190.85937 |

<div align="center">Table 14</div>

|  | Levene test | | | | Independent t-test | | | | |
|---|---|---|---|---|---|---|---|---|---|
|  | F | Significance | t | df | Significance (two tailed) | Average interpolation | SE interplation | 95% CI of interpolated Lower limit | Upper limit |
| EV | 2.526 | 0.129 | -2.017 | 18 | 0.059 | -847.26667 | 419.98559 | -1729.6237 | 35.09031 |
| Un-EV |  |  | -2.830 | 15.049 | 0.013 | -847.26667 | 299.42302 | -1485.2896 | -209.24378 |

<div align="center">Table 15</div>

|  | Number of samples | Average | SD | SEM |
|---|---|---|---|---|
| Low score group | 15 | 2153.2000 | 581.38777 | 150.11368 |
| Adult | 5 | 3780.0000 | 426.77453 | 190.85937 |

<div align="center">Table 16</div>

|  | Levene test | | | | Independent t-test | | | | |
|---|---|---|---|---|---|---|---|---|---|
|  | F | Significance | t | df | Significance (two tailed) | Average interpolation | SE interplation | 95% CI of interpolated Lower limit | Upper limit |
| EV | 1.062 | 0.316 | -5.720 | 18 | 0.000 | -1626.8000 | 284.42822 | -2224.3615 | -1029.2385 |
| Un-EV |  |  | -6.700 | 9.447 | 0.000 | -1626.8000 | 242.81972 | -2172.1625 | -1081.4375 |

<div align="center">Table 17</div>

In summary, according to the data analysis, the cross domains concept activation reaction times of the participants with high scores in basic knowledge is significantly more than that of the participants with low scores, which shows that it is difficult to activate and invoke cross domain concepts, and the reaction time data of the two groups of participants have significant statistical differences.

**7.Discussion**

7.1 Try to explore the reason

But why did this happen? It may be helpful to explain the phenomenon of this experiment that if people's thinking is patterned

Gobet and Simon (1996) tested the proficiency of Professional Chess League Champion Gary Kasparov when he playing with 4-8 chess masters at the same time. They found that despite the tremendous pressure, Kasparov performed almost as well as when he faced only one player on the tour. The two researchers believe that Kasparov's advantages are more from his ability to recognize (chess) patterns. Lesgold et al. (1988) compared the performance of five expert X-ray researchers and residents when diagnosing X-rays. They found that compared with any resident, an expert can pay more attention to the specific details on the X-ray film, give more assumptions about the causes and consequences, and combine many symptoms for analysis. However, Glaser and chi (1988) found after reviewing some studies similar to the above that experts only excelled in specialized fields. In other words, their knowledges have the characteristics of domain specificity (Kathleen M. Galotti, 2017, p181).

The patterning of thinking has also been confirmed in the field of cognitive neuroscience. Li Kuncheng and others used BOLD-fMRI to scan 9 participants when they in the task of naming pictures with different complexity (one or three words). The results showed that "the increase in the complexity of vocal naming leads to a more concentrated trend of brain activation". Li Kuncheng and others believe that this phenomenon "reflects the highly patterned of human brain in information processing" (Yang Yanhui, Lu Chunming, Li Kuncheng et al, 2008).

If it is said that people's thinking is patterned, it will help to explain the phenomenon of the result of this experiment: because people's thinking is patterned, the better the basic knowledges, the stronger the fixation of this mode, and the harder the cross mode (domain) thinking is naturally. On the contrary, if the basic knowledges is weak and the thinking mode is not established, the fetters of thinking will naturally be much less.

Of course, the above is only inference, and whether it is so still needs more targeted researches.

7.2 Divergent thinking and creativity

The experimental results of this paper easily reminiscent of the relationship between divergent thinking and creativity.

As Liu Chunlei and others put it: "it can be said that there is no very objective definition and evaluation standard for creative thinking yet" (Liu Chunlei, Wang Min, Zhang Qinglin, 2009). For a long time, there has been no clear and accepted conclusion in the debate on creativity. On its psychological basis, various schools have its own views, but the most influential ones are Wolfgang Köhler's insight learning and J.P. Guilford's divergent thinking theory. After studying the feeding behavior of the chimpanzee "Sudan", Köhler proposed that insight is a way for people or animals to reconstruct existing knowledges on solve new problems: Although Sudan knows that it can get the hanging bananas by boxes or sticks, Köhler did not teach it can use the combination of the two when the banana is too high. This method is obviously the product of its reconstruction and combination of the previous two methods (Philip G. Zimbardo, 2017, p117). Gilford's theory holds that divergent thinking is "thinking in different directions, reorganizing the information stored in the current information and memory system, and generating a large number of unique new ideas" (Liu Chunlei, Wang Min, Zhang Qinglin, 2009). As far as divergent thinking is concerned, Gilford believes that divergent thinking has the characteristics of fluency, versatility and uniqueness. It is worth noting that the "versatility" in it contains the meaning between cross domain thinking and activation (Liu Wei, 1999).

However, if the psychological basis of creativity is really divergent thinking, combined with the experimental results of this paper, we may find a contradiction: in order to make students more creative, we should focus on cultivating their divergent thinking, but in this way, we should avoid letting students master more basic knowledges. And this is obviously against common sense——almost no creation is not based on a large number of basic knowledges. Therefore, the results of this experiment may point out that there is a potential logical contradiction in the statement that divergent thinking is the psychological basis of creativity.

Of course, this paper only discusses the relationship between cross domain concept activation and the levels of their basic knowledge, not a targeted study of creativity. Therefore, the discussion in this part is only a circumstantial evidence, which is not enough to serve as a direct theoretical support.

## 8. Deficiency and improvement

Conducting research during the epidemic is the biggest challenge of this study.

Since the middle of March 2022, Beijing has started to isolation of epidemic (COVID-19). At the end of April, to comply with the requirements of the municipal government, the pupils of the whole city began to study at home. This unexpected situation brings two main problems to this study: first, the running in degree of the test questions is insufficient. In order to ensure the reliability and validity, both the basic knowledge test and the reaction time test must undergo several rounds of running in——especially the pre-test. However, students studying at home make it impossible for us to find suitable participants to carry out the pre-test. Finally, all participants underwent only one round of pre-test (the participant of the reaction time pre-test only one adult). Therefore, there is a possibility of systematic bias in the reliability and validity of the test questions in this study. The second is the problem of too few samples. At the end of June, primary schools in the city resumed classes, and the experiment could continue. But then the students faced the final examination stage soon. In order not to disturb the students' examination, we can only use the week before the summer vacation to carry out the experiment (the reaction time test is only allowed in one lesson). We tried our best, but only 22 participants were found. So the number of samples is limited. Therefore, there is a possibility that the number of participants in this study is too small and the conclusions may have biased.

The reason why the experiment should be completed in a hurry is because of the uncertainty brought by the epidemic: once large-scale isolation still occurs in the next half of the year, the research plan may be forced to be shelved for half a year again, and the previous experimental design and preliminary work may face the situation of overturn. Therefore, the title of this paper is "pre-study", it means that this paper is not a complete study. We also do not recommend the conclusion of this paper as conclusive evidence, but only as a reference.

And the conclusion of this study does not exclude another possibility: the reaction time of the high score group is larger, which is because the high score and low score students have completely different reading and semantic processing modes. There have been many previous studies on the differences of reading patterns under different cognitive levels. For example, Gao Xiaomei found in her reading study on children aged 3-6 (it is obvious that the cognitive levels of children aged 3 are different from those of children aged 6) that with the increase of age, the single visual pattern is decreasing, while the combined visual pattern is increasing (Gao Xiaomei, 2010). That is to say, there is a correlation between cognitive level and reading (specifically, this study can be analogized as reading the questions of reaction time). Then, can the reaction time still reflect the activation of concepts? Obviously, to answer this question, we need the help of eye movement research. In the future, we will try to explore this phenomenon from this perspective.

## 9. Conclusion

To sum up, the experiment shows that children's interdisciplinary activation ability of scientific concepts is negatively related to their mastery of basic knowledges. That is, the worse the basic knowledge levels, the stronger the cross domains association ability and the more flexible the thinking. We believe that the fundamental reason for this phenomenon may be that the human brain has a tendency to information processing patterning. Based on this result, this study believes that there may be a potential logical contradiction in regarding divergent thinking as the psychological basis of creativity.

As for the science teacher who triggered this study, we think the supposition cannot be excluded that his weak basic knowledge level led him associated "conductors" with "magnet".

## 10. Appendix

10.1 The basic knowledge test

Your class                                    Your name

1. Choice questions

1.1 Which is the following are not part of the POWER SOURCE: (        )

A Socket    B Bulb    C Dry cell    D 5# battery

1.2 The "?" should be connected to: (        )

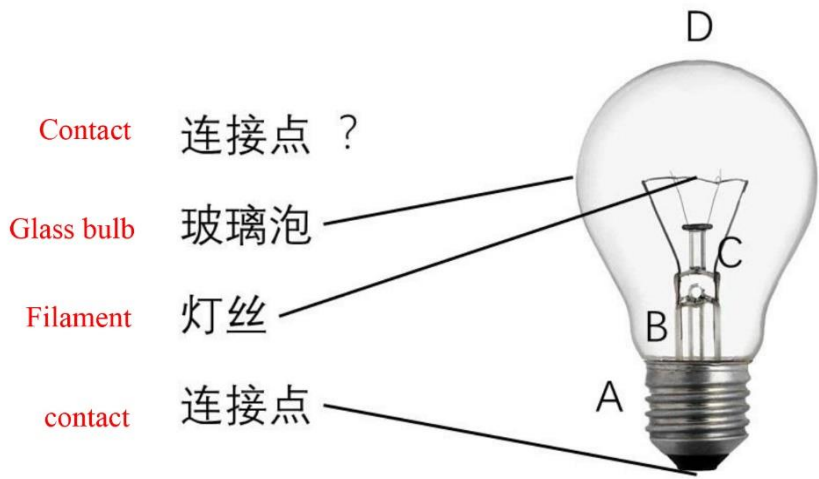

1.3 Connect the positive and negative electrodes of the battery directly without passing through any bulb or other electrical appliances: (        )

A It is a "simple circuit"

B It is a common connection method of batteries

C Attention should be paid to whether there is poor contact

D Probably will fire and explosion

1.4 Which is the following are not necessary components of a simple circuit: (        )

A Battery    B Controller    C Wireway    D Bulb

1.5 Which is the correct: (        )

A To prevent electric shock, the FIRST thing need to do before changing the bulb is to wear rubber gloves.

B If the positive and negative electrodes of the battery are reversed, the direction of the current in the circuit will also move in the opposite direction

C A simple circuit should include at least three components: battery, light bulb and switch

D Insulators are not NECESSARILY COMPLETELY non-conductive

2. The bulbs in the following circuits cannot be lit. Try to find and mark the problems (assuming all electric components are good)

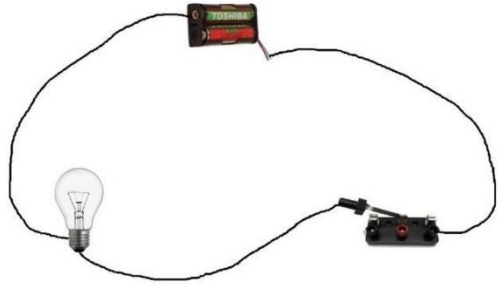

2.1

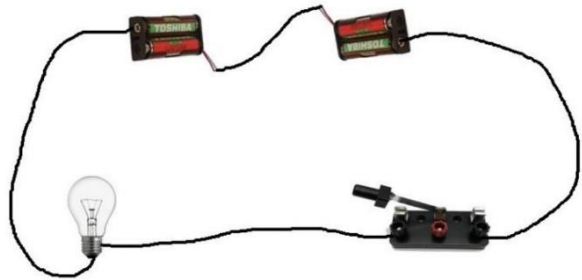

2.2

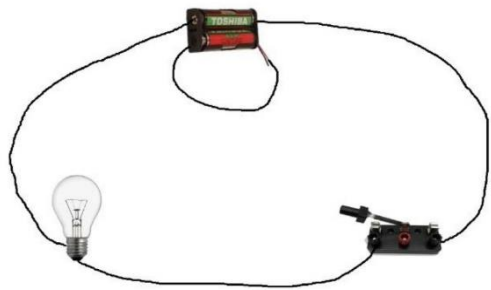

2.3

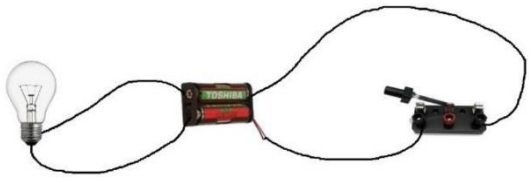

2.4

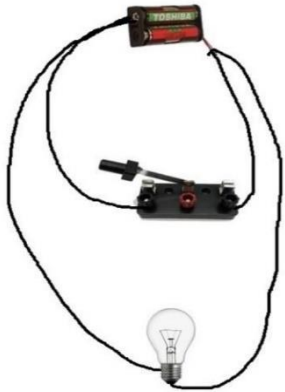

2.5

3. Try to draw the flow direction of electricity in the circuit in Fig. 2 with an arrow as shown in Fig. 1 (the arrow in Fig. 1 represents the flow direction of electricity)

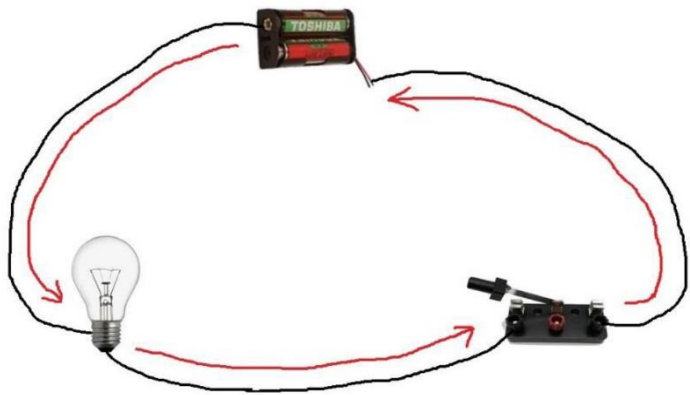

Fig.1

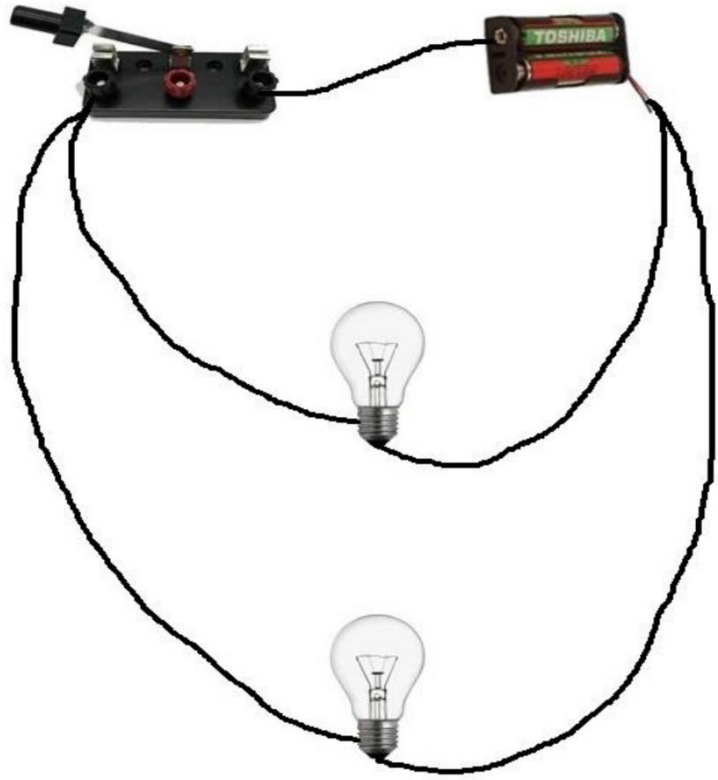

Fig.2

10.2 The data of total reaction times

| Number of participants | High/low score group | Question number | Reaction times(ms) | Accuracy |
| --- | --- | --- | --- | --- |
| 1 | High | Q1 | 2223 | 0.40 |
| | | Q2 | 1571 | |
| | | Q3 | 1367 | |
| | | Q4 | 2550 | |
| | | Q5 | 1602 | |
| | | Q6 | 2176 | |
| | | Q7 | 1654 | |
| | | Q8 | 3541 | |
| | | Q9 | 1992 | |
| | | Q10 | 2325 | |
| | | Q11 | 1751 | |
| | | Q12 | 1628 | |
| | | Q13 | 2513 | |
| | | Q14 | 2433 | |
| | | Q15 | 1594 | |
| 2 | High | Q1 | 2953 | 0.53 |
| | | Q2 | 2665 | |
| | | Q3 | 2069 | |
| | | Q4 | 3603 | |
| | | Q5 | 2723 | |
| | | Q6 | 3074 | |
| | | Q7 | 3352 | |
| | | Q8 | 4489 | |
| | | Q9 | 3139 | |
| | | Q10 | 3475 | |
| | | Q11 | 5000 | |
| | | Q12 | 2849 | |
| | | Q13 | 3698 | |

| | | Q14 | 3422 | |
| | | Q15 | 2453 | |
| 3 | High | Q1 | 2554 | 0.53 |
| | | Q2 | 4298 | |
| | | Q3 | 2654 | |
| | | Q4 | 2700 | |
| | | Q5 | 2830 | |
| | | Q6 | 3541 | |
| | | Q7 | 2669 | |
| | | Q8 | 3793 | |
| | | Q9 | 4115 | |
| | | Q10 | 3089 | |
| | | Q11 | 2803 | |
| | | Q12 | 3222 | |
| | | Q13 | 3983 | |
| | | Q14 | 3011 | |
| | | Q15 | 2171 | |
| | Average accuracy of high score group | | | 0.49 |
| 4 | Low | Q1 | 2912 | 0.60 |
| | | Q2 | 2988 | |
| | | Q3 | 1551 | |
| | | Q4 | 2892 | |
| | | Q5 | 1088 | |
| | | Q6 | 1502 | |
| | | Q7 | 1520 | |
| | | Q8 | 2267 | |
| | | Q9 | 2204 | |
| | | Q10 | 1858 | |
| | | Q11 | 2904 | |

| | | Q12 | 2189 | |
| | | Q13 | 5000 | |
| | | Q14 | 1655 | |
| | | Q15 | 2606 | |
| 5 | Low | Q1 | 1394 | 0.53 |
| | | Q2 | 1770 | |
| | | Q3 | 1198 | |
| | | Q4 | 2643 | |
| | | Q5 | 1950 | |
| | | Q6 | 1513 | |
| | | Q7 | 1513 | |
| | | Q8 | 3133 | |
| | | Q9 | 1842 | |
| | | Q10 | 1437 | |
| | | Q11 | 1187 | |
| | | Q12 | 2100 | |
| | | Q13 | 1742 | |
| | | Q14 | 1527 | |
| | | Q15 | 2633 | |
| 6 | Low | Q1 | 2634 | 0.87 |
| | | Q2 | 2861 | |
| | | Q3 | 2053 | |
| | | Q4 | 1803 | |
| | | Q5 | 2307 | |
| | | Q6 | 2029 | |
| | | Q7 | 1384 | |
| | | Q8 | 2689 | |
| | | Q9 | 3089 | |
| | | Q10 | 2368 | |

| | | Q11 | 1584 | |
| | | Q12 | 2473 | |
| | | Q13 | 2033 | |
| | | Q14 | 1844 | |
| | | Q15 | 1725 | |
| | Average accuracy of low score group | | | 0.67 |
| 7 | Adult | Q1 | 2624 | 0.67 |
| | | Q2 | 3770 | |
| | | Q3 | 3942 | |
| | | Q4 | 3000 | |
| | | Q5 | 2240 | |
| | | Q6 | 3522 | |
| | | Q7 | 2073 | |
| | | Q8 | 4389 | |
| | | Q9 | 2354 | |
| | | Q10 | 2387 | |
| | | Q11 | 2419 | |
| | | Q12 | 3255 | |
| | | Q13 | 3410 | |
| | | Q14 | 3544 | |
| | | Q15 | 2032 | |

## 10.3 The data of reaction times about cross domain concepts

| Number of participants | High/low score group | Question number | Reaction times(ms) |
| --- | --- | --- | --- |
| 1 | High | Q2 | 1571 |
| | | Q3 | 1367 |
| | | Q8 | 3541 |
| | | Q12 | 1628 |

| | | Q14 | 2433 |
|---|---|---|---|
| 2 | High | Q2 | 2665 |
| | | Q3 | 2069 |
| | | Q8 | 4489 |
| | | Q12 | 2849 |
| | | Q14 | 3422 |
| 3 | High | Q2 | 4298 |
| | | Q3 | 2654 |
| | | Q8 | 3793 |
| | | Q12 | 3222 |
| | | Q14 | 3011 |
| 4 | Low | Q2 | 2988 |
| | | Q3 | 1551 |
| | | Q8 | 2267 |
| | | Q12 | 2189 |
| | | Q14 | 1655 |
| 5 | Low | Q2 | 1770 |
| | | Q3 | 1198 |
| | | Q8 | 3133 |
| | | Q12 | 2100 |
| | | Q14 | 1527 |
| 6 | Low | Q2 | 2861 |
| | | Q3 | 2053 |
| | | Q8 | 2689 |
| | | Q12 | 2473 |
| | | Q14 | 1844 |
| 7 | Adult | Q2 | 3770 |
| | | Q3 | 3942 |
| | | Q8 | 4389 |

| | |
|---|---|
| Q12 | 3255 |
| Q14 | 3544 |

**Special thanks:** Science Teachers in Tongzhou District, Beijing, who completed the teachers' Questionnaire (see "1. The origin of this study" for details).

# Reference

Yu Bo. (2020). *Compulsory education textbooks "SCIENCE"(Volume Ⅷ)*.Educational Science Publishing House, China(Chapter 2).

Kathleen M. Galotti. (2017). *Cognitive Psychology: In and Out of the Laboratory*. China Machine Press (Chapter 7).

Philip G. Zimbardo. (2017). *Zimbardo's General Psychology*. China Machine Press (Chapter 6).

Lv Xiaojing, Ren Xuezhu. (2018,Nov). Development of associative learning and model representation of its process. *Abstracts of the 21st national psychological Academic Conference*. Beijing, BJ.

Li Qing. (2009). Research on problem representation of applied problems of children with mathematical learning difficulties in primary schools based on pass theory. *East China Normal University*,86-87

Gao Xiaomei.(2009). A study of eye movement in Chinese children's picture book reading. *East China Normal University*,3-256.

Zhang Jijia, Fu Ya, Wang Bin. (2020). Gender culture influence on spetial and weight metaphors of kinship words: Evidence from Bai, Yi, and Mosou nationalities. *Acta Psychologica Sinica*,52,440-455

Wu Yanan, Shi Lei, Qu Yunpeng et al. (2019). Two preferences of children's conceptual organization: taxonomic relationship and thematic relationship. *Psychological Research*,12(4),300-309. https://doi:10.3969/j.issn.2095-1159.2019.04.002.

Liu Chunlei, Wang Min, Zhang Qinglin. (2009). Brain mechanism of creative thinking. *Advances in Psychological Science*,17(1),106-111.

Zhang Jijia, Li Degao, Duan Xinhuan. (2010). On the conceptual connection and development of children. *Ludong University Journal (Philosophy and Social Sciences Edition)*,27(5),93-99. https://doi:10.3969/j.issn.1673-8039.2010.05.020

Yang Yanhui, Lu Chunming, Li Kuncheng et al. (2008). Neural mechanism of utterance speech expression with different complexity: a BOLD-fMRI study. *Journal of Medical Imaging*,2008(06),664. http://doi:10.3969/j.issn.1006-9011.2008.06.025.

Liu Wei. (1999). Gilford's theory and method on creative ability research. *Journal of Beijing Normal University (Social Sciences)*,1999(5),41-48.

Ye Pingzhi, Ma Qianru. (2012). Characteristics and rules of creative thinking development of children aged 2-6. *Studies in Early Childhood Education*,2012(08),36-41. http://doi:10.3969/j.issn.1007-8169.2012.08.006

Yangwen Xu, Xiaosha Wang, Xiaoying Wang et al. (2018). Doctor, Teacher, and Stethoscope: Neural Representation of Different Types of Semantic Relations. *Journal of Neuroscience.* https://doi.org/10.1523/JNEUROSCI.2562-17.2018

Sohu.com. (2020). *Lin Chongde: in the era of core literacy, where is the breakthrough in cultivating creativity?*. Retrieved from https://www.sohu.com/a/437792873_100020578/. Accessed August 24,2022.